\documentstyle[twoside,fleqn,npb,epsfig]{article}

\newcommand{\AmS}{{\protect\the\textfont2
  A\kern-.1667em\lower.5ex\hbox{M}\kern-.125emS}}

\title{Supersymmetric Sudakov corrections}

\author{M. Melles\address{Paul Scherrer Institut\\
        CH-5232 Villigen\\
        Switzerland}%
        \thanks{The author is greatful for the collaboration with M. Beccaria, F.M. Renard
	and C. Verzegnassi on the results presented.}}

\begin{document}

\begin{abstract}
For superpartner masses not much heavier than the weak scale $M=M_{\rm W}$, large logarithmic
corrections of the Sudakov type arise at TeV energies. In this paper we 
discuss the general structure of electroweak supersymmetric (susy) Sudakov corrections in the
framework of the infrared evolution equation method.
We discuss Yukawa sector Ward-identities which lead to the exponentiation of the subleading (SL)
logarithmic Yukawa enhanced Sudakov corrections in both the Standard Model (SM) as well as
in softly broken supersymmetric extensions.
The results are given to SL accuracy to all
orders in perturbation theory for arbitrary external lines in the ``light'' susy-mass scenario.
The susy-QCD limit for virtual corrections is presented.
Phenomenological applications regarding the precise determination of the important parameter $\tan \beta$
through virtual corrections
are discussed which are independent of the soft susy breaking mechanism to sub-subleading accuracy
to all orders.
\end{abstract}

\maketitle

\section{Introduction}

In light of future precision experiments in the TeV-energy regime at such machines as the LHC, TESLA or CLIC,
a lot of interest has been devoted recently to studying the high energy limit of spontaneously
broken field theories \cite{flmm,brv1,m1,m2,m3,m4,kps,kmps,dp1,dp2,ccc1,hkk,bw}. 
The main conclusion of these works, summarized in Ref. \cite{habil}
including a variety of new phenomenological applications, is that one needs to include higher order
electroweak radiative corrections through two loops at least to sub-subleading (SSL)
logarithmic accuracy \cite{kmps}. At present, only a full SL analysis in the SM to all orders
is available \cite{flmm,habil,m5,m6,kps} in the context of the infrared evolution equation
method \cite{kl}. This approach has been confirmed by explicit two loop calculations
at the leading DL \cite{m2,hkk,bw} and the SL angular dependent level \cite{dmp}.
A further SSL analysis for massless fermion production points to the necessity to include
also these SSL contributions due to large cancellations between DL, SL and SSL terms \cite{kmps}.

In general, new physics responsible for electroweak symmetry breaking is expected in the TeV regime and
the minimal supersymmetric SM (MSSM) remains an attractive candidate. If supersymmetry is relevant
to the so called hierarchy problem, then the masses $m_s$ of the new superpartners cannot be much heavier
than the weak scale $M \equiv M_{\rm W} \sim M_{\rm Z}$. In a ``light'' susy mass scenario with
$m_s \sim M$,
similarly large radiative corrections can be expected as in the SM at TeV energies. At one loop
this was confirmed by several works of the last few years \cite{brv2,brv3,brv4}.

While these corrections are of considerable phenomenological interest, it is, however, also
important to understand theoretically the high energy limit of theories which are
spontaneously broken such as the SM or the MSSM. In Ref. \cite{bmrv} an important step was
taken towards the understanding of the higher order electroweak susy Sudakov corrections.
Since the DL and angular dependent SL terms originate only from the exchange of gauge bosons,
softly broken susy does not introduce novel terms at this level. The new particle content
of the MSSM leads, however, to new universal (i.e. process independent) SL corrections
of gauge and Yukawa origin. In Ref. \cite{bmrv} it was shown that the higher order resummation
of these terms in exponential form
is a consequence of Ward-identities since also the Yukawa sector is gauge invariant.
The same reasoning lead to the exponentiation of SL Yukawa terms \cite{m3,habil}.

\begin{figure}[htb]
\vspace{9pt}
%\framebox[55mm]{\rule[-21mm]{0mm}{43mm}}
\epsfig{file=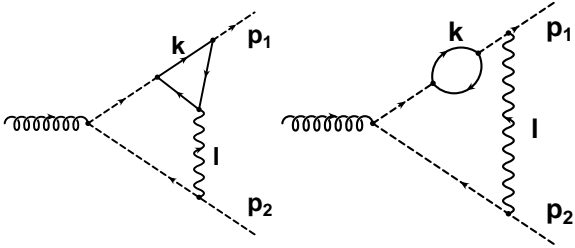,width=7.5cm}

\vspace{-0.5cm}
\caption{Two loop corrections involving Yukawa couplings
of scalars to fermions.
The Ward identity in Eq. (\ref{eq:wi}) ensures that in the Feynman gauge,
the sum
of both diagrams does not lead to
additional SL logarithms at the two
loop level.
Only corrections to the original one loop vertex
need
to be considered and lead to the exponentiation of Yukawa terms in the
SM and the MSSM to SL accuracy.} 
\label{fig:wi}
\end{figure}

\begin{figure}[t]
\vspace{9pt}
\centering
\epsfig{file=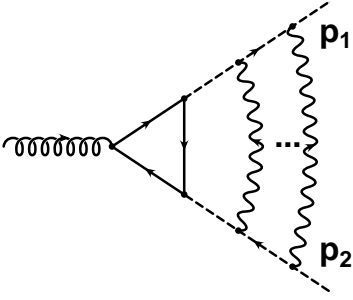,width=4.6cm}
\vspace{-0.5cm}
\caption[1]{Higher order corrections to vertices with Yukawa couplings to
SL accuracy. The graph is only schematic since in principle the gauge bosons couple
to all external legs in the process. Due to the discussion in the text the non-Abelian version
of Gribov's factorization theorem can be employed in the context of the
infrared evolution equation method.}
\label{fig:yukgt}
\end{figure}

The argument can be cast as follows.
At the two loop level, we need to
consider the diagrams displayed in Fig. \ref{fig:wi}.
The corresponding relevant parts of the two loop amplitudes read (neglecting $l$ outside the fermion loop):
\begin{eqnarray}
&&\!\!\!\!\!\!\!\!\!\!\!\!\!\! \int \!\! \frac{d^nl}{(4\pi)^n} \!\! \int \!\!
\frac{d^nk}{(4\pi)^n} \frac{
(p_1-p_2)_\nu {\rm Tr} \left[
(G_r \omega_r + G_l \omega_l) \right.}{(l^2-\lambda^2)(p_2+l)^2} \times 
\nonumber \\ &&\!\!\!\!\!\!\!\!\!\!\!\!\!\!
\frac{\left. ( \rlap/ k - \rlap/ p_1 ) 2 \rlap/ p_2 ( \rlap/ k - \rlap/ p_1
+ \rlap/ l ) (G_r \omega_r + G_l \omega_l)
 \rlap/ k \right]}{(p_1-l)^2k^2(k-p_1)^2(k-p_1+l)^2} \label{eq:ver} \\
 &&\!\!\!\!\!\!\!\!\!\!\!\!\!\! \int \!\! \frac{d^nl}{(4\pi)^n} \!\! \int \!\!
 \frac{d^nk}{(4\pi)^n} \frac{
 (p_1-p_2)_\nu {\rm Tr} \left[
 (G_r \omega_r + G_l \omega_l) \right.}{(l^2-\lambda^2)(p_2+l)^2} \times 
 \nonumber \\ &&\!\!\!\!\!\!\!\!\!\!\!\!\!\! \frac{\left.
 ( \rlap/ k - \rlap/ p_1 + \rlap/ l ) (G_r \omega_r + G_l \omega_l)
  \rlap/ k \right] 4 p_1 p_2}{(p_1-l)^2k^2(k-p_1+l)^2(p_1-l)^2}
   \label{eq:se}
   \end{eqnarray}
   where we omit common factors and the scalar masses taking $M \sim \lambda$ for clarity.
   The soft photon corrections must also be included via matching.
   The $G_{r,l}$ denote the chiral Yukawa couplings and $\omega_{r,l}=\frac{1}{2} \left(1 \pm
   \gamma_5 \right)$. The gauge coupling is written in the symmetric basis.
   For our purposes we need to
   investigate terms containing three large logarithms in those diagrams. Since the fermion
   loops at one loop only yield a single logarithm it is clear that the gauge
   boson loop momentum
   $l$ must be soft. Thus we need to show that the UV logarithm originating from the $k$
   integration is identical (up to the sign) in both diagrams.
   We can therefore neglect the loop momentum $l$ inside the fermion loop.
   We find
   for the fermion loop vertex $\Gamma^\mu(p_1^2,0,p_1^2)$ belonging to Eq. (\ref{eq:ver}):
   \begin{eqnarray}
   && \frac{1}{k^2(k-p_1)^2(k-p_1)^2} \; 
   {\rm Tr} \Big[(G_r \omega_r + G_l \omega_l)
   \nonumber \\ && ( \rlap/ k - \rlap/ p_1 ) \gamma^\mu ( \rlap/ k
   - \rlap/ p_1 ) (G_r \omega_r + G_l \omega_l)\rlap/ k \Big] \nonumber \\
   &=& \frac{4G_rG_l \left(2 p_1^\mu (k^2-p_1k) +k^\mu (p_1^2-k^2) \right)}{k^2(k-p_1)^4}
   \end{eqnarray}
   This we need to compare with the self energy loop $\Sigma (p_1^2)$ from Eq. (\ref{eq:se}):
   \begin{eqnarray}
   && \!\!\!\!\!\!\!\!\!\! \frac{\partial}{\partial {p_1}_\mu} \frac{{\rm Tr} \left[ (G_r \omega_r + G_l \omega_l)
   ( \rlap/ k - \rlap/ p_1 ) (G_r \omega_r + G_l \omega_l)  \rlap/ k \right]}{
   k^2(k-p_1)^2} \nonumber \\
   &&= \frac{\partial}{\partial {p_1}_\mu} \frac{4G_rG_l(p_1k-k^2)}{k^2(k-p_1)^2}
   \nonumber \\
   &&= 4 G_rG_l \frac{2p_1^\mu(k^2-p_1k)+k^\mu(p_1^2-k^2)}{k^2(k-p_1)^4}
   \end{eqnarray}
In short we can write
\begin{equation}
\frac{\partial}{\partial {p_1}_\mu} \Sigma (p_1^2)= \Gamma^\mu (p_1^2,0,p_1^2)
\label{eq:wi}
\end{equation}
where the full sum of all contributing self energy and vertex diagrams must be taken.
Thus, we have established a Ward identity for arbitrary Yukawa couplings of scalars to
fermions and thus, the identity of the UV singular contributions.
The relative sign is
such that the generated SL logarithms of the diagrams in Fig. \ref{fig:wi}
cancel each
other. The existence of such an identity is not surprising since it expresses the fact
that also the Yukawa sector is gauge invariant.
We are thus left with gauge boson corrections to the original vertices in the on-shell
renormalization scheme such as depicted in Fig. \ref{fig:yukgt}.
At high energies we can therefore employ the non-Abelian version
of Gribov's bremsstrahlung theorem \cite{gt}. The soft photon corrections are included via
matching as discussed in Refs. \cite{flmm,habil}. The arguments above can be applied analogously
to other external legs and lead to the exponentiation of the Yukawa and SL gauge terms
as mentioned above.

In the following we will give the complete virtual electroweak high energy corrections
for arbitrary on-shell matrix elements to all orders at SL accuracy in the ``light'' MSSM scenario.
In the general case
let us denote physical particles (fields) by $f$
and particles (fields) of the unbroken theory by $u$. Let the connection
between
them be denoted by $f=\sum_{u} C^{fu}u$, where the sum is performed over
appropriate
particles (fields) of the unbroken theory.
Note that, in general,
physical particles, having definite masses,  don't belong to
irreducible representations of the symmetry
group of the unbroken theory (for example, the photon and
$Z$ bosons have no definite isospin). On the other hand, particles of
the unbroken theory, belonging to irreducible representations of the
gauge
group, have no definite masses.
Then for the amplitude
${\cal M}^{f_1,...f_n}(\{p_k\},\{m_l\};M,\lambda)$ with $n$ physical
particles $f_i$ with momenta $p_i$ and infrared cut-off
$\lambda$, the general case for virtual corrections is given by
\begin{eqnarray}
&& \!\!\!\!\! {\cal M}^{f_1,...f_n}(\{p_k\},\{m_l\};M,\lambda) = \nonumber \\ && \!\!\!\!\! \sum_{u_1,...u_n }
\prod_{j=1}^n C^{f_ju_j}
{\cal M}^{u_1,...u_n}(\{p_k\},\{m_l\};M,\lambda) \label{eq:lc}
\end{eqnarray}
In the following we give only the corrections for a light susy mass scale $m_s \sim M$ and for
a heavy photon ($\lambda = M$) with all $|2p_lp_k| \gg m^2_s, M^2$. 
In this case, we can easily work in the symmetric
basis and give the results for these amplitudes. As discussed in Refs. \cite{flmm,habil,bmrv,bmrv2}, the
soft virtual and real QED corrections must be added by matching at the weak scale $M$.
It should be mentioned, however, that the Yukawa terms are independent of the matching terms.

Under these assumptions, we have for general on-shell matrix elements with $n$-arbitrary external
lines the following resummed SL corrections: 
\begin{eqnarray}
&& \!\!\!\!\!\!\!\!\! {\cal M}_{\rm SL}^{u_{i_1},...,
u_{i_n}} \left( \{ p_k \}; m_s ; M \right) = \exp \Bigg\{ \sum_{k=1}^n \nonumber \\ && \!\!\!\!\!\!\!\!\!
- \frac{1}{2} \left( \frac{g^2(m_s^2)}{16 \pi^2} I_k (I_k+1)
+ \frac{{g^\prime}^2(m_s^2)}{16 \pi^2} \frac{Y_k^2}{4} \right) \log^2 \frac{s}{M^2} \nonumber \\ 
&& \!\!\!\!\!\!\!\!\! 
+ \frac{1}{6} \left( \frac{g^4(m_s^2)}{64 \pi^4} I_k (I_k+1) {\tilde \beta}_0
+ \frac{{g^\prime}^4(m_s^2)}{64 \pi^4} \frac{Y_k^2}{4} {\tilde \beta}_0^\prime \right) \times \nonumber \\
&& \!\!\!\!\!\!\!\!\!
\log^3 \frac{s}{m_s^2} + \left( \delta_{i_k,B} + \delta_{i_k,{\widetilde B}} \right)
\frac{{g^\prime}^2(m_s^2)}{8 \pi^2} {\tilde \beta}^\prime_0 \log \frac{s}{m_s^2} \nonumber \\
&& \!\!\!\!\!\!\!\!\! + \left( \delta_{i_k,W^j} + \delta_{i_k,{\widetilde W}} \right)
\frac{g^2(m_s^2)}{8 \pi^2} {\tilde \beta}_0 \log \frac{s}{m_s^2} \nonumber \\
&& \!\!\!\!\!\!\!\!\!
+ \frac{g^2(m_s^2)}{16 \pi^2} C^{\rm yuk}_{i_k} \log \frac{s}{m_s^2} + 
\Bigg( \frac{g^2(m_s^2)}{16 \pi^2} I_k (I_k+1) \nonumber \\ && \!\!\!\!\!\!\!\!\! \left.
+ \frac{{g^\prime}^2(m_s^2)}{16 \pi^2} \frac{Y_k^2}{4} \Bigg)\right|_{i_k \neq \{W^j,{\widetilde W},B, 
{\widetilde B} \}} \!\!\!\!\!\!\! \log \frac{s}{M^2} \nonumber \\ 
&& \!\!\!\!\!\!\!\!\! \left.
+\frac{1}{8 \pi^2} \sum^n_{l < k} \sum_{V_a=B,W^j} \!\!\! {\tilde I}^{V_a}_{i^\prime_k,i_k}
{\tilde I}^{
{\overline V}_a}_{i^\prime_l,
i_l} \log \frac{s}{M^2} \log \frac{2 p_lp_k}{s}
\right\} \nonumber \\ &&
{\cal M}^{u_{i_1},...,u_{i^\prime_k},...,u_{i^\prime_l},...,u_{i_n}}_{\rm Born}
(\{p_{k}\})
\label{eq:angr}
\end{eqnarray}
where the index $j$ can be any value of the set  $\{1,2,3\}$.
The fields $u$ have a well defined isospin, but for angular dependent terms involving
CKM mixing effects, one has to include the extended isospin mixing appropriately in
the corresponding couplings ${\tilde I}^{V_a}_{i^\prime_k,i_k}$ of the symmetric basis.
If some of the sparticles should be heavy, additional corrections of the form
$\log^2 \frac{m_s^2}{M^2}$ etc. would be important.
Here we also assume that the asymptotic MSSM $\beta$-functions can be used with
\begin{eqnarray}
{\tilde \beta}_0&=& \frac{3}{4} C_A- \frac{n_g}{2}-\frac{n_h}{8} \\ 
{\tilde \beta}_0^\prime&=&-\frac{5}{6}n_g-\frac{n_h}{8} \label{eq:bMSSM} 
\end{eqnarray}
where $C_A=2$, $n_g=3$ and $n_h=2$. In practice, one has to use the relevant numbers of active
particles in the loops. These terms correspond to the RG-SL corrections just as in the case
of the SM as discussed in Refs. \cite{m4,ms}
but now with the MSSM particle spectrum contributing.
They originate only from RG terms within loops which without the RG contribution
would give a DL correction.

It should be noted that the one-loop RG corrections do not exponentiate and are omitted
in the above expressions. They are, however, completely determined by the renormalization group
in softly broken supersymmetric theories such as the MSSM
and sub-subleading at higher than one loop order. They can be obtained by inserting the running
one loop couplings 
\begin{eqnarray}
g^2(s) &=& \frac{g^2(m_s^2)}{1+{\tilde \beta}_0 \frac{g^2
(m_s^2)}{4\pi^2}
\ln \frac{s}{m_s^2}} \\
{g^\prime}^2 (s) &=& \frac{{g^\prime}^2 (m_s^2)}{1+{\tilde \beta}^\prime_0
\frac{{g^\prime}^2 (m_s^2)}{4\pi^2}
\ln \frac{s}{m_s^2}} \label{eq:arunMSSM}
\end{eqnarray}
into the Born cross section.
In addition, we also did not include terms discussed in Refs. \cite{m1,dp1} that are
related to the renormalization of the mixing coefficients $C^{f_ju_j}$, which in the MSSM could
be matrices.
The result in Eq. (\ref{eq:angr}), is valid for arbitrary softly broken supersymmetric extensions
of the SM with the appropriate changes in the $\beta$-functions. Taking the susy-QCD limit
($\frac{g^2}{4 \pi^2} \rightarrow \alpha_s, g^\prime \rightarrow 0, I_g (I_g+1) =
I_{\tilde g} (I_{\tilde g}+1) \rightarrow C_A=3, 
I_q(I_q+1)=I_{\tilde q} (I_{\tilde q}+1) \rightarrow C_F=4/3, 
n_h=0, M=\lambda_g= m_{\tilde g},
C^{\rm yuk}_{i_k}=0$) 
of the
various terms, Eq. (\ref{eq:angr}) is also valid for the virtual susy-QCD results. It should
be emphasized, however, that in this case the virtual corrections are not physical in the sense
that the gluon mass is zero and thus we would need to add the virtual
matching and real contributions before
we could make predictions for collider experiments, while in the SM soft QED energy cuts can 
define an observable and the heavy gauge boson masses are physical. 
In any case, the {\it form} of the operator
exponentiation in color space agrees with
the dimensionally regularized terms in Ref. \cite{cat} for non-susy QCD.

The universal SL-corrections of Yukawa type depend on the external lines only and
in the MSSM are given
by
\begin{eqnarray}
C^{\rm yuk}_{{\tilde f}_\alpha} \!\!\!\!\!&=&\!\!\!\!\! - 
\frac{1}{4} \Bigg( \left[ 1+\delta_{\alpha,{\rm R}} \right] \frac{{\hat m}
^2_{\tilde f}}{M^2} + \delta_{\alpha,{\rm L}}
\frac{{\hat m}^2_{{\tilde f}^\prime}}{ M^2} \Bigg)
\nonumber \\ \!\!\!\!\!&=&\!\!\!\!\! C^{\rm yuk}_{f_\alpha} \label{eq:fs}
\\ C^{\rm yuk}_{\phi^\pm} \!\!\!\!\!&=&\!\!\!\!\! - \frac{3}{4} \Bigg( \frac{m_t^2}{M^2} + \frac{m_b^2}{M^2}
\Bigg) = C^{\rm yuk}_\chi \! = C^{\rm yuk}_{H_{\rm SM}} \label{eq:ps} \\ 
C^{\rm yuk}_{H^\pm} \!\!\!\!\!&=&\!\!\!\!\! 
- \frac{3}{4} \Bigg( \frac{m_t^2}{M^2} \cot^2 \beta + \frac{m_b^2}{M^2}
\tan^2 \beta \! \Bigg) \!\!= C^{\rm yuk}_A \label{eq:Hs} \\ 
C^{\rm yuk}_{{\tilde H}_\alpha} \!\!\!\!\!&=&\!\!\!\!\!
- \frac{3}{4} \Bigg( \frac{m_t^2}{M^2} \left[ 1 + \cot^2 \beta \right]
\delta_{\alpha,\rm L}  \nonumber \\ \!\!\!\!\!&&\!\!\!\!\! 
+  \frac{m_b^2}{M^2} \left[ 1 + \tan^2 \beta \right]\delta_{\alpha,\rm R} \Bigg)
\label{eq:Hos}
\end{eqnarray}
The sfermion chiralities ($\alpha$) are those of the fermions ($f$) whose superpartner is produced.
In addition we denote ${\hat m}_{\tilde f}=m_t / \sin \beta$
if ${\tilde f}= {\tilde t}$ and ${\hat m}_{\tilde f}=m_b / \cos \beta$ if ${\tilde f}={\tilde b}$.
${\tilde f}^\prime$ denotes the corresponding
isopartner of ${\tilde f}$. For particles other than those belonging to the third family
of quarks/squarks, the Yukawa terms are negligible.
Eq. (\ref{eq:fs}) displays
an exact supersymmetry in the sense that the same corrections are obtained for the fermionic and sfermionic
sector in the regime above the electroweak scale $M$.
The same holds also for the remaining relative SL corrections from the full result in Eq. (\ref{eq:angr})
to all orders. Since we assume $m_s\sim M$, we omit mixing effects which could be important for larger
mass gaps.

Note the additional factor of $3=N_C$ in Eqs. (\ref{eq:Hs}), (\ref{eq:ps}) and (\ref{eq:Hos}) compared
to Eq. (\ref{eq:fs}), 
leading to a significant dependence on the important parameter $\tan \beta = \frac{v_u}{v_d}$,
the ratio of the two vacuum expectation values.  
More importantly, terms depending on soft breaking parameters like mass ratios enter only at the
sub-subleading (SSL) level. In Refs. \cite{brv3,bmrv} this point was emphasized and used for a determination
of $\tan \beta$, based on the above virtual electroweak corrections, through a one loop subtraction method,
which at SSL level is independent of the soft susy breaking terms.  
At higher orders, there are terms of ${\cal O} \left( \alpha^n C_{\rm soft} \log^{2n-2} \frac{s}{M^2} \right)$,
however, if the condition $s \sim s^\prime \gg M^2$ is fulfilled, the difference
in the cross section measurements at $s$ and $s^\prime$ will be proportional
to ${\cal O} \left( \alpha^n C_{\rm soft} \log \frac{s}{s^\prime} \log^{2n-3} \frac{s}{M^2} \right)$
which is of beyond the SSL approximation.

This model independence is a clear and important difference to other ways of measuring $\tan \beta$ like 
${\tilde \tau}$-decays \cite{baer,den,hin} and should be utilized at future TeV linear colliders.
The possible relative precision depends strongly on the accuracy of the measured cross sections.
Assuming 10 one percent measurements between 0.8 and 3 TeV the precision
ranges from 
50 \% for $\tan \beta \geq 10$,
25 \% for $\tan \beta \geq 15$ and a few percent measurement for
$\tan \beta \geq 25$.

In conclusion, we have presented fully general virtual SL results to all orders in the context of the
MSSM for arbitrary on-shell matrix elements. 
The form of these corrections can be written in exponential operator form in the $n$-particle space
in the symmetric basis in the light susy scenario. Subleading universal terms exponentiate due to
Ward identities while angular dependent corrections are determined by Born-rotated matrix elements
in complete analogy to the SM.

The size of these contributions depends crucially on
the energy of future colliders and it is clear that two 
(at CLIC possibly three) loop electroweak corrections cannot be omitted
at machines operating at TeV energies if the desired precision is at the percentile level. 
A precise determination of $\tan \beta$, independent of the soft breaking terms to SSL accuracy, should be 
utilized at future linear colliders. Work towards completing this program on the theoretical
side is in progress \cite{bmrv3}.

\end{document}